# Pouzyry: a novel class of algorithms for restoring a function from a random sample

F.V.Tkachov
*Institute for Nuclear Research of Russian Academy of Sciences
Moscow, 117312, Russian Federation*

A novel class of algorithms for restoring a function from a random sample is based on the concept of weak convergence, borrows algorithmic solutions from the Optimal Jet Finder (hep-ph/0301185), offers a considerable algorithmic flexibility, is applicable to non-positive functions, is insensitive to the choice of coordinate axes. A first implementation demonstrates feasibility of the approach.

*Small random raindrops
can hardly hurt pouzyry.
Oh! joy of research*

The fundamental problem of modeling a function from a random sample has at least two important applications: multi-dimensional adaptive MC integration (for a review see [1]) and construction of quasi-optimal observables for data analysis [2].

The conventional view is that a function is a way to provide a number $f(x)$ for any number $x$. However, with $x$ measured via a finite precision measurement procedure, increasing the number of (unbiased, independent) measurements increases the precision of the estimate of $x$ (by taking the standard average) — but for $f$ one only obtains the average $\int f(x)\varphi(x)dx$ where $\varphi$ is the probability distribution for individual measurements of $x$. By improving the measurement procedure one makes $\varphi$ more narrow and thus can approach $\int f(x)\varphi(x)dx \to f(x_0)$ — but only if $f$ is continuous at $x_0$. Otherwise the result depends on the shape of $\varphi$. However, in practice one rarely if ever cares about how the function is defined at the points of discontinuity.

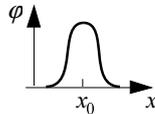

Therefore, it is logically sufficient to define a function by its averages $\langle f,\varphi\rangle = \int f(x)\varphi(x)dx$ with all possible test functions $\varphi$ that possess continuous derivatives of any order (this is a technical restriction imposed for technical convenience without loss of meaning) and are equal to zero outside bounded regions (each $\varphi$ has its own such region). The averages $\langle f,\varphi\rangle$ are linear in $\varphi$, and one can define a "generalized function" as an arbitrary linear correspondence $f:\varphi \to \langle f,\varphi\rangle$; a convenient abuse of notation is to write $\langle f,\varphi\rangle = \int f(x)\varphi(x)dx$. A familiar example is Dirac's $\delta$-function.

To discuss approximations, one must specify the meaning of the proposition that a sequence of (generalized) functions $f_n$ converges to a (generalized) function $f$. The convergence motivated by the conventional definition of functions is the pointwise convergence, i.e. a convergence (with uncorrelated rates) of all numerical sequences $f_n(x) \to f(x)$ for all $x$. Within the framework of the new interpretation, the true argument of a function is not $x$ but $\varphi$, and the notion of convergence is modified accordingly: $\langle f_n,\varphi\rangle \to \langle f,\varphi\rangle$ for all $\varphi$, with uncorrelated rates. In practice $n$ stays finite, and one ensures a smallness of the differences $\langle f_n,\varphi\rangle - \langle f,\varphi\rangle$ for a finite set of $\varphi = \varphi_k$.

We write $f_n \xrightarrow[n\to\infty]{\text{weak}} f$ and use the qualifier "weak" to describe this type of convergence and related notions (closeness, etc.). As a first heuristic approximation, one may rely on the analogy between the weak convergence and metric convergences.

There are many advantages in replacing the archaic

"general functions", i.e. mappings $x \to f(x)$, with the subtler notion of "generalized functions", i.e. linear mappings $\varphi \to \langle f, \varphi \rangle$, in our mental arsenal of mathematical concepts [3]. It is remarkable that such a finesse of interpretation results in truly powerful new options for constructive problem solving. In the context of particle physics, one example is the long-sought solution of the problem of asymptotic expansions of Feynman diagrams [4], which required an essential use of techniques of generalized functions (see a discussion in [5]). Another example is the discovery of the optimal jet definition [6] where the optimal configuration of jets is regarded as an approximation in the sense of weak convergence. It turns out that the latter idea has a much wider range of applicability, as is shown below.

Consider a random sample of values $\{x_n\}_{n=1}^N$ of a random variable $x$ distributed according to $\pi(x) \geq 0$. The opening idea of the theory of statistics is that for $N \to \infty$, the sample reproduces the probability distribution $\pi(x)$. What would be a precise interpretation for that? More generally, given a random sample $F_N = \{x_n, f_n\}_{n=1}^N$, where $f_n = f(x_n)$, what's the precise meaning of the statement that $F_N$ represent $f(x)\pi(x)$ increasingly well for $N \to \infty$?

Represent $F_N$ as a sum of $\delta$-functions: $F_N(x) \equiv N^{-1} \sum_n f_n \delta(x - x_n)$. Then the required precise interpretation is as follows:

$$F_N(x) \xrightarrow[N \to \infty]{\text{weak}} f(x)\pi(x), \qquad (1)$$

i.e. for any test function $\varphi$, the sequence of its integrals with the l.h.s. converges to its integral with the r.h.s., $\int f(x)\pi(x)\varphi(x)dx$, in the usual sense.

I am not aware of a textbook that would state this in an explicit fashion. This may be explained by the fact that mathematical statistics had already matured [7] by the time the ideas of generalized functions only began to be publicised [8]. It is important to clearly understand, however, that *the interpretation (1) is an essential starting point for all the thinking* about how to obtain more tractable approximations for the r.h.s. instead of the l.h.s.; the latter, however, is the only possible starting point (perhaps, along with some a priori information about $f\pi$ that can be used to fine-tune the algorithms).

We will call such constructive approximations *models*, generically denote them as $M(x)$, and require that a model $M(x)$ provide constructive algorithms for:

A) the mapping $x \to M(x)$ for real $x$;

B) generation of random $x$ distributed according to $M(x)$, provided the latter is non-negative.

So, one has to find a model $M_{N,f}(x)$ that would be close in the weak sense to the "raw" approximation $F_N(x)$, which fact would guarantee that $M_{N,f}(x)$ remains close to $f(x)\pi(x)$ in the weak sense:

$$\frac{1}{N} \sum_n f_n \delta(x - x_n) \bullet \xrightarrow[N \to \infty]{\text{weak}} \bullet f(x)\pi(x)$$
$$\searrow_{\text{weak}} \bullet M_{N,f}(x)$$

Moreover, if we impose restrictions on $M_{N,f}(x)$ in accordance with whatever a priori information we may have about $M_{N,f}(x)$, we may hope that $M_{N,f}(x)$ is close to $M_{N,f}(x)$ in a stronger sense (uniform, etc.). Mathematical results of this type are well known [9].

In practice, the following types of models are used:

**(i) Decompositional models:** the function's domain of definition $\mathcal{D}$ is split into non-intersecting subdomains, $\mathcal{D} = \bigcup_k \mathcal{D}_k$, $\mathcal{D}_k \cap \mathcal{D}_{k'} = \varnothing$, and the model is defined to be constant in each subdomain: $M(x)|_{\mathcal{D}_k} = \text{const}$. $\mathcal{D}_k$ can either be fixed (as with standard histograms) or found adaptively.

**(ii) Galiorkin models:** $M(x) = \sum_k m_k \mathcal{P}_k(x)$, where $\mathcal{P}_k(x)$ is an orthogonal system of functions, and $m_k = \int dx\, F_N(x) \mathcal{P}_k(x)$.

**(iii) Parametric models:** one chooses a function parameterized by a number of parameters and adjusts the latter to fit the sample.

**(iv) The Vegas model** is employed in the Vegas routine for multidimensional integration [10]. It is a direct product of one-dimensional adaptive decompositional models, $M(x) = \prod_i M_i(x_i)$. The popularity of Vegas shows that even such very crude approximation can be a valuable model in many dimensions.

**(v) Kernel models.** Let $K(x)$ be any convenient (usually hat-like) function such that

$$K_R(x) \equiv R^{-\dim} K(R^{-1}x) \xrightarrow[R \to 0]{\text{weak}} \delta(x). \qquad (2)$$

Then it is sufficient to replace the individual $\delta$-functions $\delta(x-x_n)$ in $F_N(x)$ with $K_R(x-x_n)$. In general, $R$ should be smaller for larger $N$.

**(vi) NN models.** The most popular simplest neural networks [11] are described by the analytical expression $M(x) = g\left(\sum_k c_k g\left(\sum_j A_{kj} x_j + B_k\right)\right)$, where $g$ is a smooth step-like function. If one drops the outermost $g$ (which is irrelevant in the present context), there remains a linear combination of rotated and shifted step-like functions. This is to be compared with the kernel models: rotated and shifted step-like functions roughly correspond to infinite-$R$ kernels positioned at infinite points in various directions.

The kernel models (v) remain, perhaps, least studied. The approach seems to become impractical for large $N$ — the case which is often the most interesting. It could be advantageous to "condense" the sum of $\delta$-functions to a **(i)** smaller number of **(ii)** more regularly distrubuted $\delta$'s. This must be done so as to ensure a weak closeness of the condensed sum to the original one:

$$F_N(x) \stackrel{weak}{\approx} \frac{1}{P}\sum_p \tilde{f}_p \delta(x-\tilde{x}_p) + \tilde{f}_0 \equiv \tilde{F}_P(x), \qquad (3)$$

i.e. so as to minimize the differences

$$\left|\langle F_N \varphi\rangle - \langle \tilde{F}_P \varphi\rangle\right| \qquad (4)$$

for test functions $\varphi$. Then the scheme becomes

$$\frac{1}{N}\sum_n f_n \delta(x-x_n) \bullet \xrightarrow[N\to\infty]{weak} \bullet f(x)\pi(x)$$
$$\frac{1}{P}\sum_p \tilde{f}_p \delta(x-\tilde{x}_p) + \tilde{f}_0 \bullet \longrightarrow \bullet \frac{1}{P}\sum_p \tilde{f}_p K_R(x-\tilde{x}_p)$$
$$(5)$$

The replacement (3) based on minimization of (4) is exactly what is effected in the optimal jet definition [6]. Repeating the reasoning of [6] with appropriate simple modifications, one arrives at the following criteria for finding $\tilde{x}_p$ and $\tilde{f}_p$:

One introduces an $N \times P$ matrix, $0 \le z_{n,p} \le 1$, $\bar{z}_n \equiv 1 - \sum_p z_{n,p} \ge 0$, and sets $\tilde{f}_p = \sum_n z_{n,p} f_n$, $\tilde{f}_p \tilde{x}_p = \sum_n z_{n,p} f_n x_n$, so that $z_{n,p}$ becomes the unknown. The matrix $z$ is found from the requirement of minimization of the following expression:

$$\Omega_R = \frac{1}{R^2}\sum_{p,n} z_{n,p}|f_p|(x_n-\tilde{x}_p)^2 + \sum_n \bar{z}_n|f_p|.$$

This controls the remainder of a Taylor expansion of (4) in $x_n - \tilde{x}_p$, with the leading terms nullified by the above restrictions on $\tilde{x}_p$ and $\tilde{f}_p$. The parameter $R$ remains free.

The minima of $\Omega_R$ correspond to configurations with $z_{,p}$ equal to 0 or 1. The set of sample points with $z_{n,p}=1$ constitutes the $p$-th *pouzyr*[1]. $\tilde{x}_p$ and $\tilde{f}_p$ are the pouzyr's location and weight (or charge). The domain spanned by a pouzyr is always convex, centered at $\tilde{x}_p$, and its radius does not exceed $R$.

It should be empasized that the criterion of minimizing $\Omega_R$ is a constructive expression of the requirement to make the original configuration of sample points and the resulting configuration of pouzyry as close as possible in the weak sense.

It is remarkable that the first step of the model construction within the pouzyry approach (the left downward arrow in (5)) is performed within the realm of singular generalized functions. In this respect the pouzyry scheme is entirely novel and unusual.

An exploratory algorithm to minimize $\Omega_R$ was obtained by modifying the Optimal Jet Finder [12]. To ensure mathematical correctness of the resulting modified algorithm and that subtle bugs are not introduced during the modification, I chose to work with the statically safe, modular, object-oriented programming language Component Pascal [13], and chose the verification version of the optimal jet finder [14], designed for robustness, as a starting point for modifications.

I report the following initial findings:
1) One has to choose $P$ and $R$ prior to running the minimization. From general considerations, $P$ should not be chosen larger than $O(\sqrt{N})$ (which is large enough for practical purposes). An optimal choice of $R$ depends on the function; it may e.g. correspond to a typical distance over which first order derivatives of the function vary appreciably.
2) The key element of the minimum search algorithm here is an iteration step which scans all sample points once and modifies the configuration of pouzyry. This iteration step is the only highly technical element of the algorithm not really subject to variations (apart from possible optimizations). All other elements allow variations. For instance, the shapes of kernels are arbitrary. Similarly, each kernel in the resulting sum can be given its own $R$ depending on the effective radius of the corresponding pouzyr. Such variations should be employed to incorporate the properties of the solution to maximal degree.

---
[1] Pouzyr means "bubble" in Russian, the plural being pouzyry, to be pronounced according to the rules of French.

3) The time required to execute a single iteration step is the same as in the case of the Optimal Jet Finder [12], $O(P \times N)$, i.e. linear in both the number of sample points and the number of pouzyry. The CPU time per iteration is a fraction of a second on a 866 MHz computer for dim = 7, $N = 200$, $P = 10$.

4) The resulting configuration of pouzyry depends on the initial one (the starting point for minimum search). With a purely random choice (all $z_{n,p}$ random) all the pouzyry are initially located near the middle of the integration domain (the effect of averaging). This may not be optimal. It may help to devise smarter ways to choose the initial configuration.

5) In some cases (e.g. for large $R$) one may observe the following behavior: at first the configuration of pouzyry converges pretty fast, but then the convergence slows down greatly; the configuration may change significantly over $O(100)$ iterations. This means that a straightforward minimization may not be an optimal strategy in the more complex cases.

6) Several ideas to improve upon the simplest pouzyry scheme suggest themselves, namely: (a) to "breed" better configurations of pouzyry using e.g. the ideas of genetic algorithms; (b) to make $R$ depend on $p$ in the formula for $\Omega_R$; (c) to seek a model which is a sum of pouzyry configurations with various $R$: contributions with larger $R$ would describe larger-scale behavior of the function, whereas contributions with smaller $R$ would describe narrow structures.

7) Since OJF reliably finds narrow clusters [12], the pouzyry scheme is guaranteed to reliably find narrow spikes in the initial sample. More generally, the better a narrow structure can be approximated by a sum of spikes, the better pouzyry would work.

8) As an example, consider the function $f(x) = \text{MAX}(1 - r/\rho, 0)$, where $r$ is the euclidean distance from $x$ to the diagonal of the hypercube, and $\rho$ describes the width of the diagonal strip within which $f$ is non-zero. For dim = 7, $\rho = 0.1$, $N = 200$ (only sample points with non-zero values of $f$ were retained and counted), $P = 10$, $R = 0.1$, it takes about 15 iterations to reach a minimum [15]. The quality of the resulting model can be judged from how much better is the MC integration with the probability distribution given by the model compared with the simplest MC with uniformly distributed sample points. In the present case, despite the obviously too low $P$ (not enough to cover the entire diagonal at the chosen $R$), the statistical integration error is improved by a factor of 7 which translates into a factor of 49 reduction of CPU time. Note that the Vegas algorithm is defeated by such diagonal structures, whereas the pouzyry scheme is insensitive to the choice of coordinates. On the other hand, with an unfortunate choice of $P$ or $R$, the pouzyry scheme may not yield any significant improvement over the simplest Monte Carlo integration.

9) Pouzyry, which deals well with narrow structures, could be combined with other methods well suited for description of the smooth global structure of the function, e.g. the Galiorkin methods. Note that the pouzyry scheme does contain a constant background already ($f_0$ in (3))

The described pouzyry scheme is novel and rather unusual, and given all the algorithmic options it opens, it is hard to assess its potential at present. I should be content to have demonstrated its feasibility.

I thank A.Czarnecki for the hospitality at the University of Alberta (Edmonton, Canada) where the idea of *pouzyry* came to me; V.Ilyin for comments; K.Tobimatsu for encouragement. This work was supported in parts by the Natural Sciences and Engineering Research Council of Canada, KEK, the organizing committee of the ACAT'03 workshop, and the CERN Theory Division where the final version was prepared.